\documentclass[final,5p,times,twocolun]{elsarticle}
\usepackage{lineno} %hyperref
\usepackage{amsmath}           %allows the use of align
\usepackage{amssymb}
\usepackage{bm}                %bold in math mode

\usepackage{graphicx}
\usepackage[space]{grffile}    %allows spaces in picture names

\usepackage{tabularx}
\usepackage{todonotes}
\usepackage[separate-uncertainty,multi-part-units = single, allow-number-unit-breaks]{siunitx}
\usepackage{subfig}
\usepackage{url}

\modulolinenumbers[5]

\journal{NIM A}

\newcommand{\arco}{$\textrm{Ar}\textnormal{-}\textrm{CO}_2$}
\newcommand{\arcois}[1]{$\textrm{Ar}\textnormal{-}\textrm{CO}_2$ (#1)}
\newcommand{\neco}{$\textrm{Ne}\textnormal{-}\textrm{CO}_2$}
\newcommand{\necois}[1]{$\textrm{Ne}\textnormal{-}\textrm{CO}_2$ (#1)}
\newcommand{\baseline}{$\textrm{Ne}\textnormal{-}\textrm{CO}_{2}\textnormal{-}\textrm{N}_{2}$}

\newcommand{\E}[1]{$\textrm{E}_{\textrm{#1}}$}

\newcommand{\ti}[1]{$t_{\textrm{#1}}$}

\begin{document}

\begin{frontmatter}

\title{Ion mobility measurements in \arco{}, \neco{}, and \baseline{} mixtures, and the effect of water contents}

\author[GSI,HD]{Alexander Deisting\corref{corraut}}
\cortext[corraut]{Corresponding author}
\ead{alexander.deisting@cern.ch}
\author[GSI]{Chilo Garabatos}
\author[Bratislava]{Alexander Szabo}

\address[GSI]{GSI Helmholtzzentrum f\"ur Schwerionenforschung GmbH, Planckstra{\ss}e 1, Darmstadt, Germany}
\address[HD]{Physikalisches Institut, Ruprecht-Karls-Universit\"at Heidelberg, Heidelberg, Germany}
\address[Bratislava]{Faculty of Mathematics, Physics and Informatics, Comenius University, Bratislava, Slovakia}

\begin{abstract}
A detector has been constructed for measuring ion mobilities of gas mixtures at atmospheric pressure and room temperature. The detector consists of a standard triple GEM amplification region and a drift region where ions drift. A method has been developed to measure the ions' arrival time at a cathode wire-grid by differentiating the recorded signals on this electrode. Simulations prove that this method is accurate and robust. The ion mobility in different gas mixtures is measured while applying different drift field values ranging from \SI{200}{\volt\per\centi\meter} to \SI{1100}{\volt\per\centi\meter}.\\ \indent
From an extrapolation of a Blanc's law fit to measurements in \arco{} mixtures we find the reduced mobility of the drifting (cluster) ion species in pure argon to be $\SI{1.94(1)}{\centi\meter\squared\per\volt\per\second}$ and in pure carbon-dioxide to be $\SI{1.10(1)}{\centi\meter\squared\per\volt\per\second}$. Applying the same procedure to our measurements in \neco{} yields $\SI{4.06(7)}{\centi\meter\squared\per\volt\per\second}$ and $\SI{1.09(1)}{\centi\meter\squared\per\volt\per\second}$ for the reduced mobilities in pure neon and carbon-dioxide, respectively.\\ \indent
Admixtures of $\textrm{N}_2$ to \necois{90-10} reduce somewhat the mobility. For the baseline gas mixture of the future ALICE Time Projection Chamber, \baseline{} (90-10-5), the measured reduced mobility of the drifting ions is \SI{2.92(4)}{\centi\meter\squared\per\volt\per\second}.\\ \indent
Ion mobilities are examined for different water content ranging from \SI{70}{ppm} to about \SI{2000}{ppm} in the gas using \arcois{90-10} and \necois{90-10}.  A slight decrease of ion mobility is observed for the addition of several hundred ppm of water.
\end{abstract}


\end{frontmatter}

%\linenumbers

\section{Introduction}
The ion mobility ($K$) is the factor relating the drift velocity ($\textrm{v}_{\textrm{Drift}}$) of ions to the drift field (\E{Drift}): $\textrm{v}_{\textrm{Drift}} = K \cdot \textrm{E}_{\textrm{Drift}}$ \cite{mcdaniel1973mobility}. Under high particle loads, \textit{i.e.} on the order of several $\si{\nano\ampere\per\square\centi\meter}$ reaching the readout plane, ions can accumulate in the drift volume of a gaseous detector. These ions produce a large space charge distorting the drift field. Such distortions and their possible mitigation have been and are studied by many experiments as \textit{e.g.} the current ALICE Time Projection Chamber \cite{alme2010alice} (TPC) \cite{distortionNote}, the upgraded ALICE TPC \cite{bohmer2013simulation,aliceTpcUpgradeTDR2014} and the LCTPC \cite{krautscheid2015design,1748-0221-13-04-P04012}. If $K$ is known, it is possible to estimate and simulate this space charge and to evaluate the impact of the back drifting ions on the performance of a detector.\\ \indent
Mobilities of ions in their parent gases (\textit{e.g.} $\textrm{Ne}^{+}$ in neon) have been measured extensively, \textit{cf.} \cite{Ellis1976,Ellis1978,Ellis1984,VIEHLAND199537}, but the mobility of an ion species in a gas mixture differs from its mobility in the parent gas. With this study we provide ion mobilities for different gas mixtures commonly used in gaseous detectors and in particular we measure the mobility in the gas mixture of the upgraded ALICE TPC, \baseline{} (90-10-5) \cite{aliceTpcUpgradeTDR2014}.\\ \indent
To measure the ion drift velocity a dedicated set-up is constructed, which is described in Section \ref{ch:immeas::sec:setupsketch}. The principle of the measurements and the signal analysis procedure follow in Section \ref{ch:immeas:sec:ideaOfmeasurement}. In Section \ref{ch:immeas:subsec:tGrid} we describe our method to determine the ions' time of arrival at the end of the drift gap and we demonstrate with simulations that the method is accurate and robust. Afterwards the measurement of the signal indicating that the ions started to drift (Sec. \ref{ch:immeas:subsec:tGEM}) and the different uncertainties affecting the mobility measurements (Sec. \ref{ch:immeas:subsec:allerrors}) are discussed. The ion mobility found for different \arco{} and \neco{} gas mixtures are presented as a function of \E{Drift} (Sec. \ref{ch:immeas:subsec:koversusE}) and as function of the quencher content in the gas mixture (Sec. \ref{ch:immeas:subsec:blancarconeco}). Furthermore the effect of admixtures of $\textrm{N}_2$ to \necois{90-10} is examined (Sec. \ref{ch:immeas:subsec:addingn2}) and the effect of water on the ion mobility is studied using \arcois{90-10} and \necois{90-10} (Sec. \ref{ch:immeas:subsec:resultswater}). We summarise our findings in Section \ref{ch:immeas:sec:summary}.

\section{Experimental set-up}
\label{ch:immeas::sec:setupsketch}
\begin{figure}%l u r 0
\centering
\includegraphics[width=\columnwidth,trim = 0 0 0 0, clip=true]{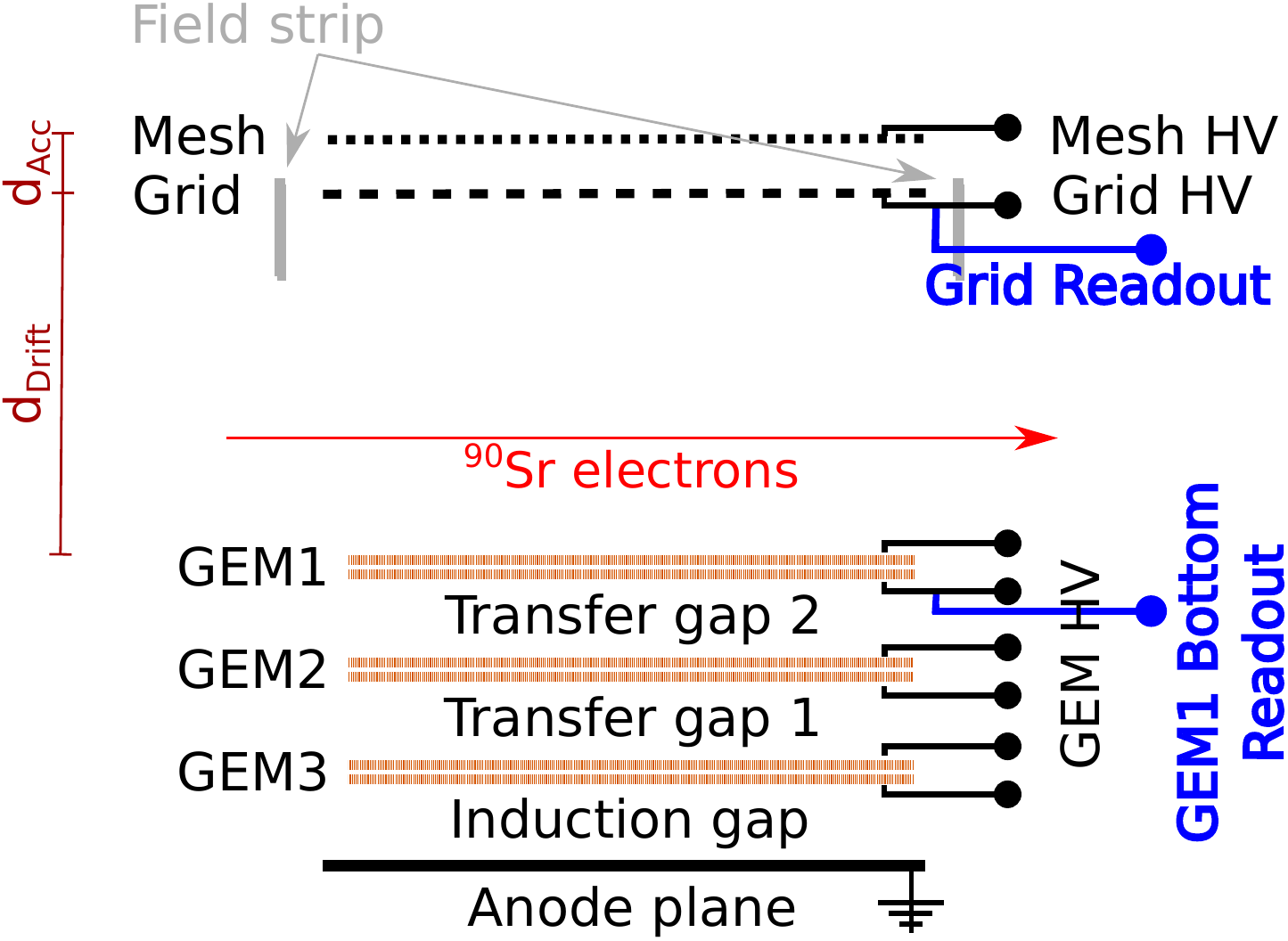}
\caption{\label{ch:immeas:fig:setupsketch}Set-up for the measurement of ion mobility. Signals from the GEM1 bottom electrode and the wire-grid are passed on to a preamplifier and to a digital oscilloscope where individual signals are stored for later processing or several signals are averaged and then the average is stored.}
\end{figure}

The experimental set-up is described in \cite{Deisting2017215}, nevertheless we give a short summary. We use a detector (Fig. \ref{ch:immeas:fig:setupsketch}) with a triple Gas Electron Multiplier (GEM) \cite{sauli1997gem} stack with standard $10\times\SI{10}{\centi\meter\squared}$ foils. The transfer gaps 1 and 2, and the induction gap have a width of \SI{2}{\milli\meter}. Above the GEM stack a wire-grid (wire spacing/diameter: \SI{2}{\milli\meter}/\SI{100}{\micro\meter}) is mounted at a distance of $d_{\textrm{Drift}}=\SI{21.35(12)}{\milli\meter}$. The grid is followed at a distance $d_{\textrm{T3}}=\SI{6.4(2)}{\milli\meter}$ by a mesh serving as drift cathode. This region between grid and mesh (GEM1 top, respectively) is referred as \textit{transfer gap 3} (\textit{drift gap}, respectively). A field strip mounted at the GEM-stack-side of the grid is used to improve the field homogeneity.\\ \indent
We use an open gas system employing Bronkhorst F-201CV \cite{mfc} mass flow controllers which allow one to mix up to three different gases. The detector exhaust consist of several metres of \SI{4}{\milli\meter} pipe. The temperature is monitored by the lab's air conditioning system and the atmospheric pressure is recorded by a transmitter of the CTE7000 series \cite{CTE7000} located elsewhere. Both values are used to correct the measured mobilities. A DewMaster water sensor \cite{waterSensor} is placed shortly after the detector in the gas exhaust line, to monitor the water content of the counting gas. The sensor is a double-cooling stage chilled mirror hygrometer with a precision of \SI{1}{ppm}.\\ \indent
Each electrode is powered with an individual channel of either a CAEN N471 \cite{2chps} Power Supply (PS) or a CAEN N470 \cite{4chps} PS, therefore, the voltage on each electrode can be tuned individually.\\ \indent
Signals are decoupled from the High Voltage (HV) line to the wire-grid using a $C_{\textrm{Dec}}=\SI{20}{\nano\farad}$ capacitor and afterwards the signals are fed into an ORTEC 142IH \cite{ORTEC142IH} preamplifier. A similar preamplifier is used to decouple and process the signals from the HV line supplying the GEM1 bottom electrode. This preamplifier is suitable for millisecond rise time signals, as confirmed with a pulse generator.

\section{Principle of measurement}
\label{ch:immeas:sec:ideaOfmeasurement}
\begin{figure*}
\centering%l u r 0
\subfloat[]{\includegraphics[width=\columnwidth, trim = 126 0 106 55, clip=true]{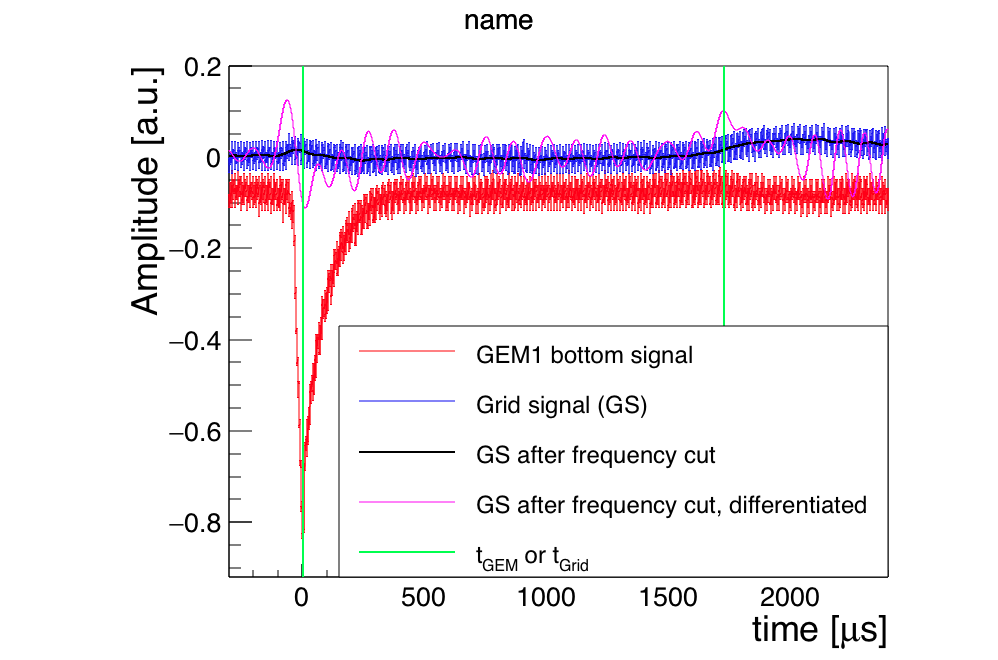}
\label{ch:immeas:fig:singleEvent}}
\subfloat[]{\includegraphics[width=\columnwidth, trim = 126 0 106 55, clip=true]{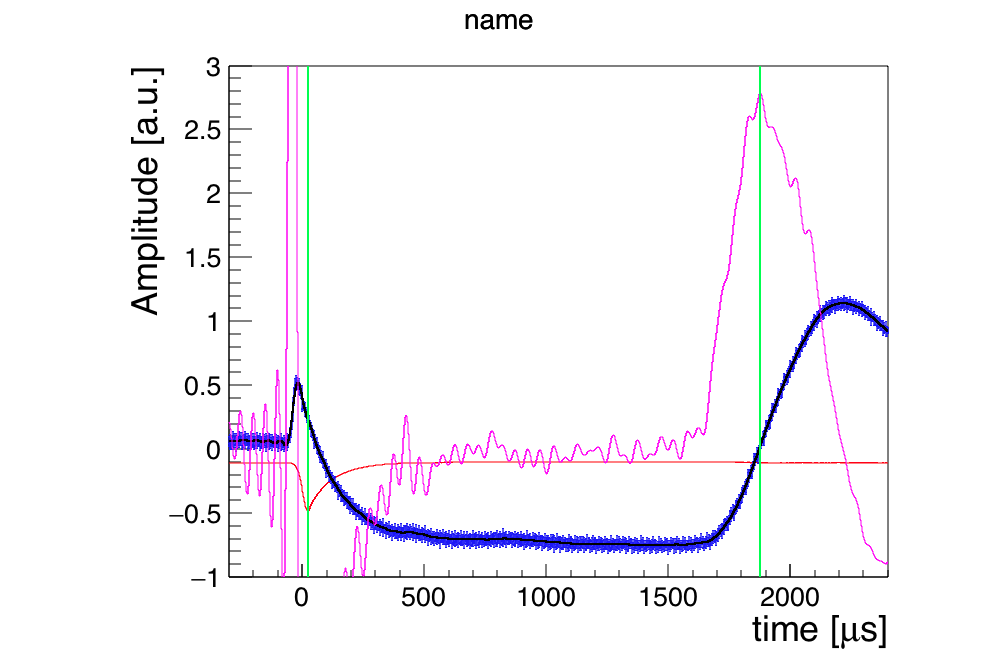}
\label{ch:immeas:fig:averEventfull}}
\caption{\label{ch:immeas:fig:IMtestEvents}\protect\subref{ch:immeas:fig:singleEvent} A simultaneous measurement of the signal from the bottom side of GEM1 and the wire-grid and \protect\subref{ch:immeas:fig:averEventfull} the average of 2000 of such measurements. The legend in Figure \protect\subref{ch:immeas:fig:singleEvent} applies to both plots. In \protect\subref{ch:immeas:fig:averEventfull} the \textit{Grid signal} is scaled up by a factor 100 and is shifted at the vertical axis, to fit the scaled signal again into the plot. All measurements are done in \necois{90-10}, but the arrival times differ because the gas pressure and temperature in \protect\subref{ch:immeas:fig:singleEvent} and \protect\subref{ch:immeas:fig:averEventfull} differ.}
\end{figure*}
Electrons from primary tracks in the drift gap drift to the GEM stack. The ions from the gas amplification produced in GEM3 and GEM2 drift through the stack and induce a signal on the GEM1 bottom electrode, as they leave the GEM stack at a time $t_{\textrm{GEM}}$. Afterwards the ions drift in a uniform electric field towards the wire-grid. Behind the grid a similar field as in the drift gap is set. Throughout the entire time of ions moving in the drift gap and in the transfer gap 3, a signal is induced on the grid. The time at which the ions cross the grid ($t_{\textrm{Grid}}$) is extracted from this signal (Sec. \ref{ch:immeas:subsec:tGrid}). We use the time difference $t_{\textrm{Drift}} = t_{\textrm{Grid}} - t_{\textrm{GEM}}$ and the known length of the drift gap to calculate the ions velocity $\textrm{v}_{\textrm{Drift}} = \frac{d_{\textrm{Drift}}}{t_{\textrm{Drift}}}$. Then the ion mobility is calculated from this velocity and the voltage difference across the drift gap $\Delta U_{\textrm{Drift}}=U_{\textrm{GEM1\ Top}}-U_{\textrm{Grid}}$
\begin{align}
\begin{split}
K &= \frac{\textrm{v}_{\textrm{Drift}}}{\textrm{E}_{\textrm{Drift}}} = \frac{\textrm{v}_{\textrm{Drift}}\cdot d_{\textrm{Drift}}}{\Delta U_{\textrm{Drift}}} = \frac{(d_{\textrm{Drift}})^2}{t_{\textrm{Drift}}\cdot\Delta U_{\textrm{Drift}}} \quad .
\end{split}
\label{ch:immeas:eq:KfromMeas}
\end{align}
\label{ch:immeas:sec:dataana}\label{ch:immeas:subsec:FFT}Because the signals are affected by a regular noise in the form of oscillations in the frequency range between about \SI{40}{\kilo\hertz} to about \SI{200}{\kilo\hertz} (\textit{e.g.} Fig. \ref{ch:immeas:fig:singleEvent}), an average of 2000 subsequent signals is used for the data analysis. The average signal is transformed, through fast Fourier transform, into the frequency domain, where frequencies above $f_{\textrm{Cut}}=\SI{10}{\kilo\hertz}$ are cut out and then the signals are transformed back. Figure \ref{ch:immeas:fig:IMtestEvents} illustrates the difference between an individual signal (Fig. \ref{ch:immeas:fig:singleEvent}), the average over 2000 signals (Fig. \ref{ch:immeas:fig:averEventfull}) and the corresponding signals after the frequency cut has been applied.

\subsection{Measurement of the ion arrival at the grid}
\label{ch:immeas:subsec:tGrid}
\begin{figure}[h]
\centering%l u r 0
\includegraphics[width=\columnwidth, trim = 62 0 53 18, clip=true]{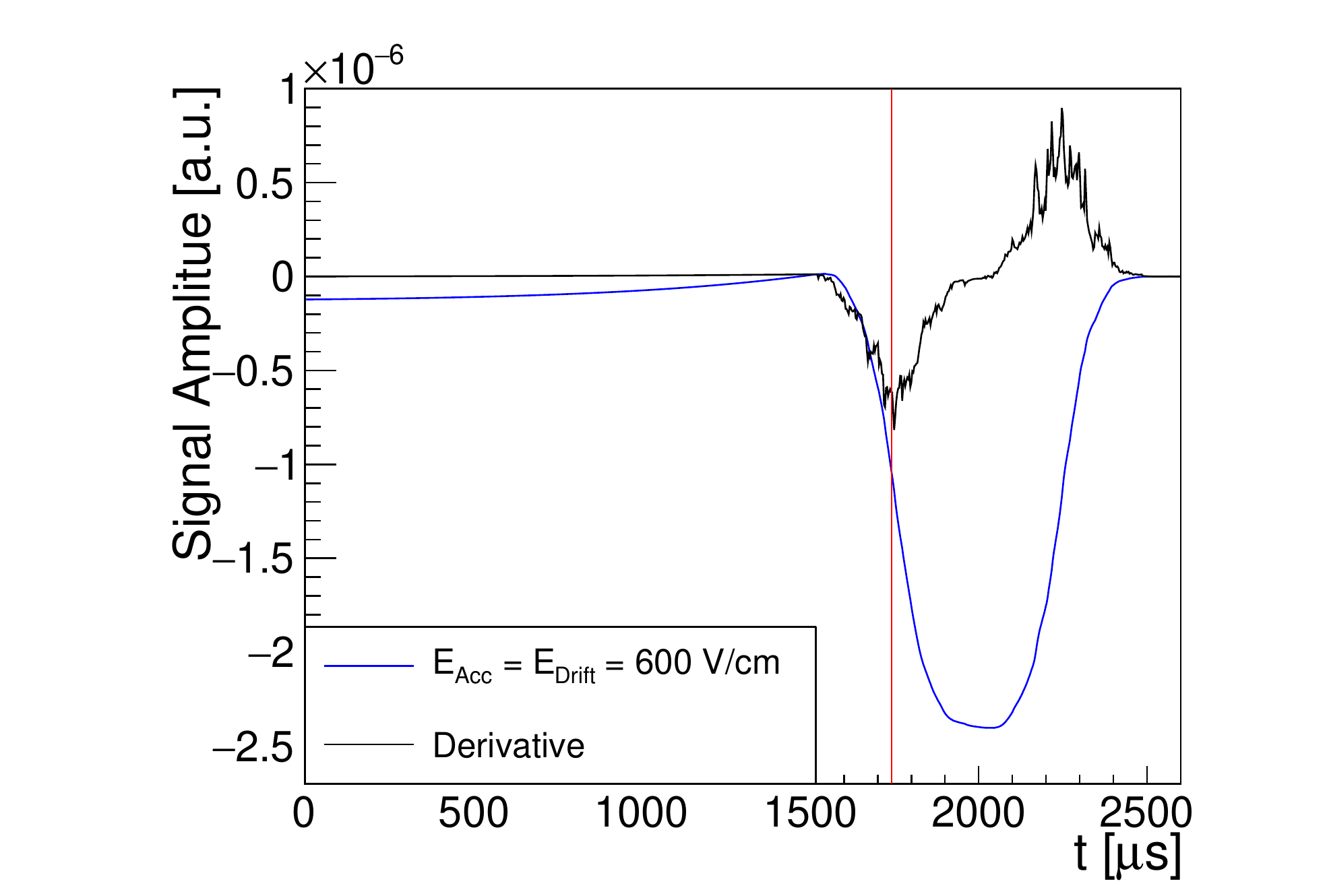}
\caption{\label{ch:immeas:fig:edriftconsteaccdiffnormandderivative}Garfield simulation of the induced ion signal on a wire-grid and its derivative. The vertical red line marks the time at which the ions are expected to arrive at the grid ($t_{\textrm{Grid}}^{\textrm{Expected}}$).}
\end{figure}

\begin{figure*} 
\centering%l u r 0
\subfloat[]{\includegraphics[width=\columnwidth, trim = 110 0 92 55, clip=true]{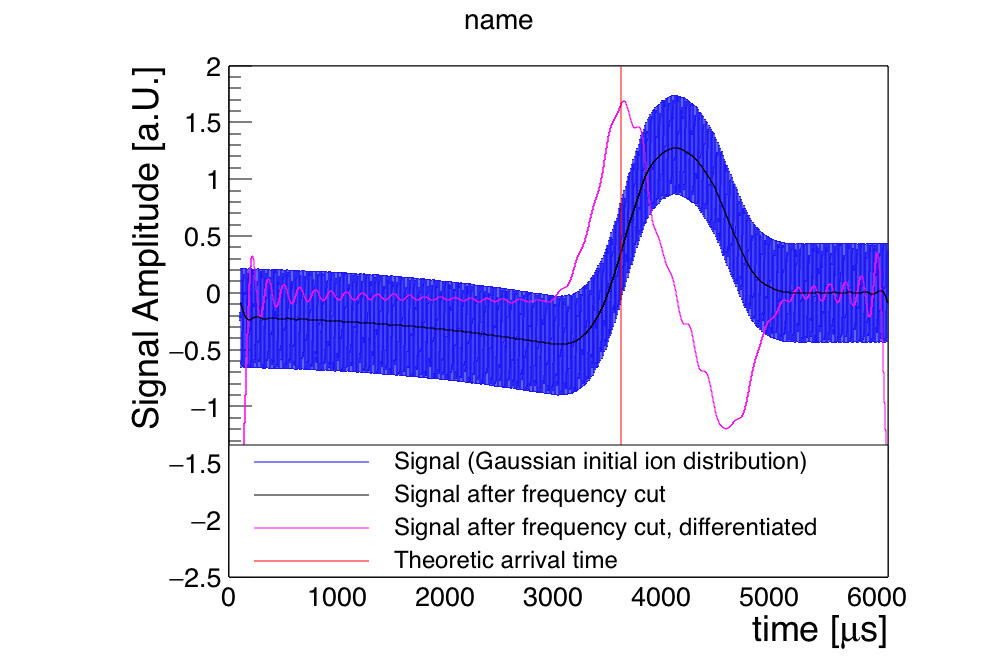}
\label{ch:immeas:fig:simEventAna}}
\subfloat[]{\includegraphics[width=\columnwidth, trim = 110 0 92 55, clip=true]{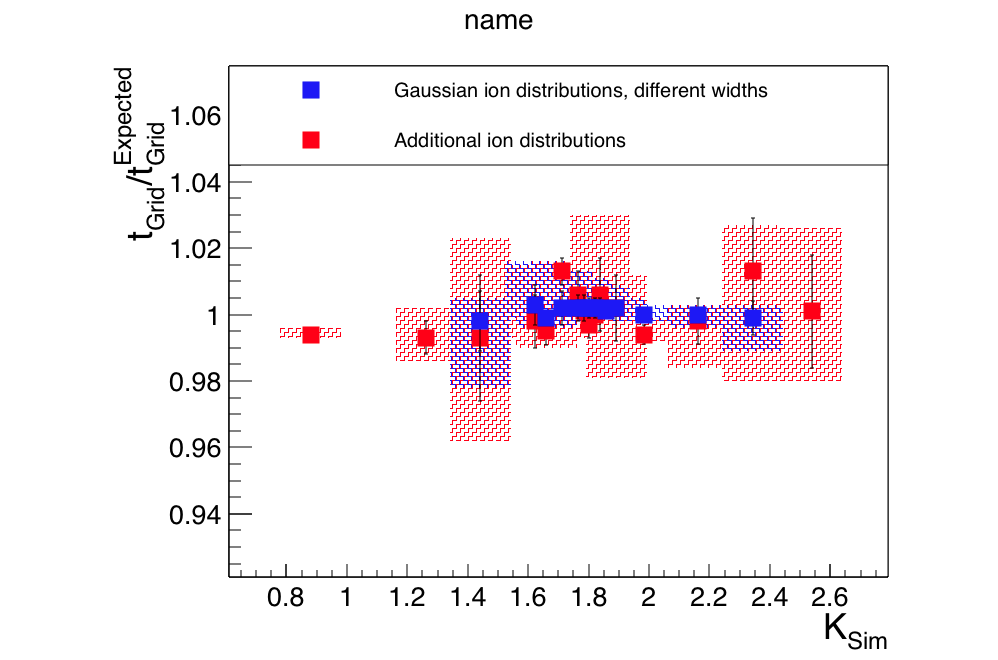}
\label{ch:immeas:fig:tresultsdifferentk}}
\caption{\label{ch:immeas:fig:anachaintest}\protect\subref{ch:immeas:fig:simEventAna} With Garfield the signal induced by ions on the wire-grid is simulated and overlaid with a periodic noise. These modified simulations are analysed with the procedure used for measured signals. Signal simulations, as in Figure \protect\subref{ch:immeas:fig:simEventAna}, are done for varying initial ion distribution and mobilities ($K_{\textrm{Sim}}$). Different threshold frequencies are used for the frequency cut during the signal analysis procedure. The time at which the ions pass the wire-grid (\ti{Grid}) is extracted from these signals and compared to the value expected from the input parameters of the simulation ($t_{\textrm{Grid}}^{\textrm{Expected}}$). \protect\subref{ch:immeas:fig:tresultsdifferentk} $t_{\textrm{Grid}}/t_{\textrm{Grid}}^{\textrm{Expected}}$ as function of $K_{\textrm{Sim}}$. The vertical bands show the full range of different \ti{Grid} extracted for different $f_{\textrm{Cut}}$, while the average (respectively standard deviation) of the corresponding \ti{Grid} values is marked by a square (respectively the error bars).}
\end{figure*}

At the time the ions pass through the grid, the polarity of the induced signal changes \cite{ramo1939currents}. In the two plots in Figure \ref{ch:immeas:fig:IMtestEvents} this is visible in the \textit{Grid signal}.\footnote{The preamplifiers invert the actual signals.} The \textit{Differentiated grid signal} in Figure \ref{ch:immeas:fig:averEventfull} has distinct peaks, which allow to identify the inflection points. The last inflection point corresponds to the time when ions cross the wire-grid, \textit{i.e.} \ti{Grid}, if the electric fields in the drift gap and transfer gap 3 are the same. Simulations explained below show the robustness of this method.\\ \indent
A configuration with a wire-grid and an homogeneous electric field \E{Drift} (\E{T3}, respectively) in the drift gap (transfer gap 3, respectively) is used to generate signals with Garfield \cite{veenhof1984garfield} as ions drift and cross the grid. An arbitrary ion mobility is introduced into the simulations and $10^5$ ions are initially distributed such that they cover several wire pitches. In drift direction the ion distribution has a Gaussian shape. The drifting ions' signal on the wire-grid is shown in Figure \ref{ch:immeas:fig:edriftconsteaccdiffnormandderivative} for ions drifting first through the drift gap and then through the transfer gap 3. When the ions approach the grid the signal rises and it becomes more negative after the ions crossed the grid. Eventually the signal is zero, when the ions reach the cathode. Knowledge of \E{Drift}, the inserted ion mobility and the drift distance yields the expected time of arrival at the grid, $t_{\textrm{Grid}}^{\textrm{Expected}}$. However, the signal shape lacks a distinctive feature, which is suitable to measure the time when ions pass through the grid.\\ \indent
Figure \ref{ch:immeas:fig:edriftconsteaccdiffnormandderivative} shows as well the signals' derivative. There is an inflection point during the period when the signal becomes more negative, which is visible as a peak. This inflection point occurs at $t_{\textrm{Grid}}^{\textrm{Expected}}$, if \E{T3}$\ =\ $\E{Drift}. Therefore the inflection point in the ion signal on the wire-grid is suitable to determine $t_{\textrm{Grid}}$, if the electric field on both sides of the grid is the same. \\ \indent
\label{ch:understandingTimesBetter}In order to test the reliability of this inflection point method the simulations are tuned to reproduce the shape of measured signals including the overlaying of a periodic noise (Fig. \ref{ch:immeas:fig:simEventAna}). The resulting waveforms are then fed in the data analysis chain and the ratio of $t_{\textrm{Grid}}^{\textrm{Expected}}$ to \ti{Grid} is determined. This is done for different mobilities, varying \E{Drift} and \E{T3}, and different initial ion distributions. For each simulation several values of $f_{\textrm{Cut}}$ are used during the analysis. The range of electric fields in the simulation is similar to the actual fields used during the measurements. Furthermore, the pressure (temperature) in the simulation is set to atmospheric pressure (room temperature) and varied in order to cover the gas conditions during our measurements. The results in Figure \ref{ch:immeas:fig:tresultsdifferentk} illustrates that the inflection point of the grid signal is indeed a reliable measure to determine \ti{Grid} if \E{Drift}$\ =\ $\E{T3}. The additional ion distributions in the figure include square, triangular, and asymmetric multiple-Gaussian distributions. Figure \ref{ch:immeas:fig:tresultsdifferentk} shows that the results for $t_{\textrm{Grid}}^{\textrm{Expected}}/$\ti{Grid} do not depend on the initial ion distribution in the simulation as long as the drift distance is measured from the centre of gravity of the ion distribution to the grid.

\subsection{Discrimination between ions with different mobilities}
Simulations similar to the ones presented in Figure \ref{ch:immeas:fig:anachaintest} are used to examine the case where several ion species with different mobilities drift. The induced signals on the wire grid is simulated in a first step for the respective single ion species. Afterwards, these signals are overlaid to yield one signal of several ions with different mobility. Proceeding in steps is necessary, because Garfield does not offer the possibility to introduce more than one ion mobility. If the arrival time of two ion species differs by more than $\Delta$\ti{Grid}$=\SI{100}{\micro\second}$, distinct inflection points are obtained for each species. Using Equation \eqref{ch:immeas:eq:KfromMeas}, $\Delta$\ti{Grid} can be translated to a relative difference between two mobilities, which can be resolved by our method:
\begin{align*} 
\frac{\Delta K}{K} = \frac{\Delta t_{\textrm{Grid}} \cdot \textrm{E}_{\textrm{Drift}}}{d_{\textrm{Drift}}} \cdot K \quad ,
\end{align*}
\textit{e.g.} for $d_{\textrm{Drift}}=\SI{21.53}{\milli\meter}$, $\textrm{E}_{\textrm{Drift}}=\SI{400}{\volt\per\centi\meter}$, and $K=\SI{2}{\centi\meter\squared\per\volt\per\second}$ $\Delta K / K$ is \SI{3.7}{\%}.

\subsection{Ion signal at the GEM1 bottom electrode}
In the simulations discussed in Section \ref{ch:understandingTimesBetter} the initial ion distribution is an input to the calculations, so its centre of gravity is known as is the start-time $t_{\textrm{GEM}}$ of the ion mobility measurement. In the real set-up the ion distribution is a result of the gas amplification in the GEM stack. The distribution is measured using the induced ion signal at the GEM1 bottom electrode. Ions from the electron amplification in GEM3 dominate the signal at our high voltage settings, however, there is a small contribution of ions originating at GEM2. These ions are spread in space, hence $t_{\textrm{GEM}}$ should be defined as the time when half of the ions has crossed GEM1 and entered the drift gap.

\subsection{Measurement of the ions entering the drift gap}
\label{ch:immeas:subsec:tGEM}
The signal induced on the GEM1 bottom electrode rises as the ions move towards it, and falls when the ions either cross GEM1 or are collected by the electrode (Fig. \ref{ch:immeas:fig:IMtestEvents}). The maximum of this signal thus corresponds to the time when the ions reach the level of GEM1 bottom. The voltage settings of the GEM stack are tuned to reach a small width of the ion signal's peak. Ideally \ti{GEM} is the time at which the ions pass the GEM1 top electrode, hence an offset due to the width of the GEM foil ($\sim\SI{50}{\micro\meter}$) is introduced using the peak in the GEM1 bottom signal. The time it takes one ion to cross GEM1 can be estimated to be lower than \SI{1}{\micro\second}, using conservative values for the mobility and the electric field inside the GEM hole ($K=\SI{1}{\centi\meter\squared\per\volt\per\second}$, \E{GEM}$=\SI{10}{\kilo\volt\per\centi\meter}$). On the other hand, the field leakage from the GEM1 holes into the drift volume produces a small, local increase of the drift field near the GEM. We therefore ignore these two, competing effects and chose the start-time of the ions' drift time measurement ($t_{\textrm{GEM}}$) to be the time of the highest amplitude in the GEM1 bottom signal. 

\section{Results}
\label{ch:immeas:sec:results}
For a given gas mixture, the induced signals at the GEM1 bottom electrode and at the wire-grid are simultaneously recorded. These signals are averaged over 2000 or 3000 subsequent events for each field setting. The ions' drift time is extracted as explained in Section \ref{ch:immeas:subsec:tGrid} and \ref{ch:immeas:subsec:tGEM} and the ion mobility $K(\textrm{E}_{\textrm{Drift}})$ is calculated (Eq. \eqref{ch:immeas:eq:KfromMeas}). We apply the commonly used density correction (\textit{e.g.} \cite{mcdaniel1973mobility,Ellis1976}) to the gas density at \SI{273.15}{\kelvin} and \SI{1013}{\milli\bar} to calculate the reduced mobility $K_{0}$ using $T_{\textrm{Meas}}$ and $P_{\textrm{Meas}}$ during each measurement.\footnote{The pressure and temperature in our detector is on average $T_{\textrm{Meas}}\sim\SI{293.5}{\kelvin}$ and $P_{\textrm{Meas}}\sim\SI{966}{\milli\bar}$ with a standard deviation of \SI{1.5}{\kelvin} and \SI{9}{\milli\bar}, respectively.}
\begin{align}
K_0 = K \cdot \frac{\SI{273.15}{\kelvin}}{T_{\textrm{Meas}}} \cdot  \frac{P_{\textrm{Meas}}}{\SI{1013}{\milli\bar}}
\label{ch:immeas:reducedMobility}
\end{align}
The actual results are shown from Section \ref{ch:immeas:subsec:koversusE} on. In the following we discuss shortly the uncertainties affecting our measurement.

\subsection{Estimate of uncertainties in the experiment}
\label{ch:immeas:subsec:allerrors}
\subsubsection{Time measurements}
\label{ch:immeas:subsec:timeerror}
The drift time of the ions ($t_{\textrm{Drift}}$ in Eq. \eqref{ch:immeas:eq:KfromMeas}) is calculated from \ti{GEM} and \ti{Grid}. The start time, \ti{GEM}, is determined as the time at which the peak in the signal from the GEM1 bottom electrode reaches its most negative amplitude (\textit{cf.} Sec. \ref{ch:immeas:subsec:tGEM}). In Figure \ref{ch:immeas:fig:IMtestEvents} \ti{GEM} is marked by the first vertical line. A minimum (maximum, respectively) of the oscillations peaking close to (at, respectively) the actual minimum in the signal, will lead to a shift of the measured \ti{GEM}. Therefore the uncertainty on \ti{GEM} is half an oscillation period, \textit{i.e.} $\delta(t_{\textrm{GEM}})=\SI{20}{\micro\second}$.\\ \indent
The \ti{Grid} measurement is examined with the simulations discussed in Section \ref{ch:immeas:subsec:tGrid}. The removal of the oscillations on top of the actual signal introduces an uncertainty on \ti{Grid}. After applying the frequency cut, oscillations remain in the grid signal and in turn in its derivative. The location of the maxima in the derivative's oscillations influences the found \ti{Grid}. We quantified the uncertainty on the \ti{Grid} measurement due to this effect with simulations of the wire-grid signal (Sec. \ref{ch:immeas:subsec:tGrid}). Figure \ref{ch:immeas:fig:tresultsdifferentk} shows the ratio $t_{\textrm{Grid}}/t_{\textrm{Grid}}^{\textrm{Expected}}$ and the spread of this quantity occurring when $f_{\textrm{Cut}}$ is varied. Based on this test we conclude that the relative error of our method to determine \ti{Grid} is not larger than $\pm\SI{2}{\%}$. The error on \ti{Grid} is hence $\delta(t_{\textrm{Grid}}) = 0.02 \cdot t_{\textrm{Grid}}$.\\ \indent
Gaussian error propagation is used to calculate the error on the full drift time $t_{\textrm{Drift}} = t_{\textrm{Grid}} - t_{\textrm{GEM}}$
\begin{align}
\begin{split}
\delta\left(t_{\textrm{Drift}} \right) &= \sqrt{\delta(t_{\textrm{Grid}})^2 + \delta(t_{\textrm{GEM}})^2} \\
 &= \sqrt{(0.02\cdot t_{\textrm{Grid}})^2 + (\SI{20}{\micro\second})^2} \quad .
\end{split}
\label{ch:immeas:eq:fullTimeError}
\end{align}

\subsubsection{Set-voltages}
\label{ch:immeas:subsec:othererrors}
The uncertainty on each set voltage is $\delta(U_{\textrm{Set}})=\SI{2}{\volt}$. Therefore the full uncertainty on the voltage difference across the drift gap is
\begin{align}
\begin{split}
\delta&(\Delta U_{\textrm{Drift}})=\\
&\sqrt{\delta(U_{\textrm{GEM1\ Top}})^2+\delta(U_{\textrm{Grid}})^2} = \sqrt{2} \cdot \SI{2}{\volt} \quad .
\end{split}
\label{ch:immeas:eq:Udrifterror}
\end{align}
This uncertainty is fully correlated among subsequent measurements without changing the voltages in-between. However, the mobility is always measured for different \E{Drift} and therefore different voltage settings.

\subsubsection{Drift length}
The error on the drift length $\delta(d_{\textrm{Drift}})$ is a systematic uncertainty, which is fully correlated among all measurements with the same $d_{\textrm{Drift}}$. The measurement of the spacers used to define the drift length yields an uncertainty of about \SI{0.1}{\milli\meter} (Sec. \ref{ch:immeas::sec:setupsketch}).

\subsubsection{Gas conditions}
We assume conservative statistical errors of the pressure and temperature of $\delta(P_{\textrm{Meas}})=\SI{1}{\milli\bar}$ and $\delta(T_{\textrm{Meas}})=\SI{0.5}{\kelvin}$, respectively. The flow stability of the employed mass flow meters is \SI{0.1}{\%}. Considering the highest and lowest flow used during all measurements, the largest uncertainty on a mixing ratio is estimated to be lower than \SI{1.7}{\%}. \textit{E.g}: The uncertainty of a \SI{10}{\%} component of a gas mixture is thus \SI{0.17}{\%}. 

\begin{figure*}
\centering%l u r 0
\subfloat[\arco{} mixtures]{\includegraphics[width=\columnwidth,trim = 63 0 45 30, clip=true]{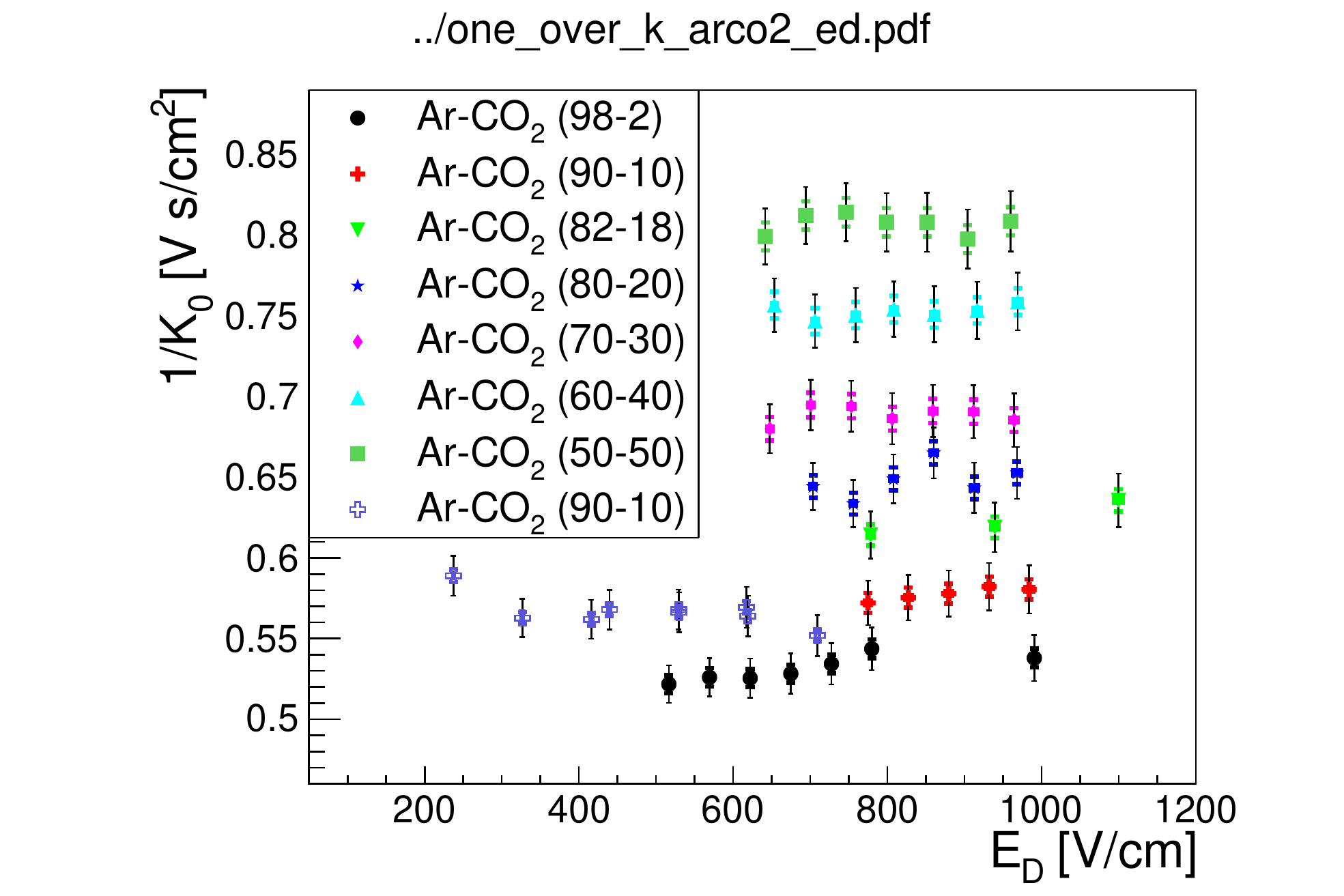}\label{ch:immeas:fig:result:ArCO2ED}}
\subfloat[\neco{} mixtures]{\includegraphics[width=\columnwidth,trim = 63 0 45 30, clip=true]{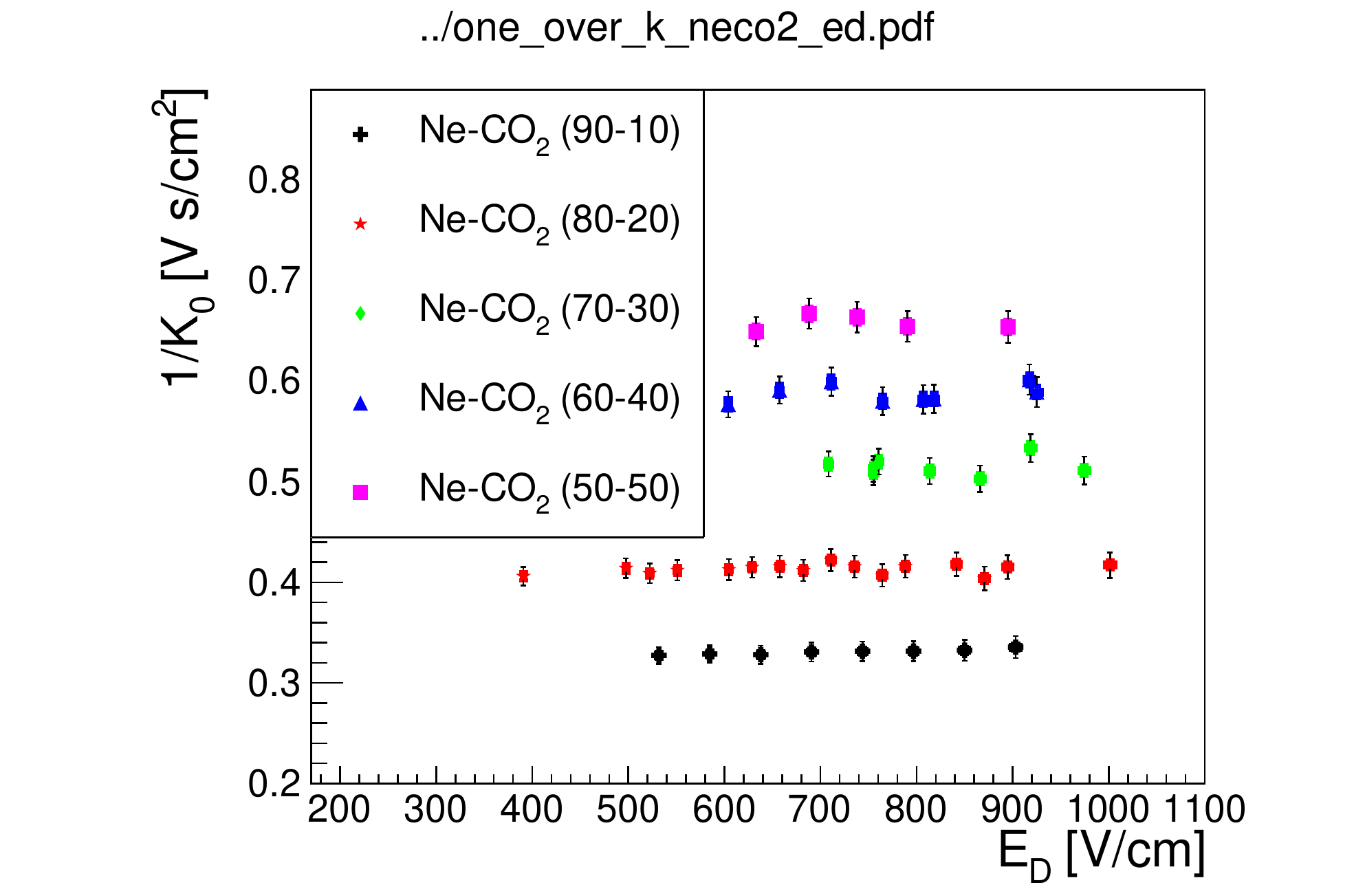}\label{ch:immeas:fig:result:NeCO2ED}}
\caption{\label{ch:immeas:fig:result:ArAndNeCO2ED}Inverse reduced ion mobility for \arco{} (\neco{}, respectively) mixtures. All closed (open, respectively) points are measured with a drift length of \SI{21.35}{\milli\meter} (\SI{25.31}{\milli\meter}, respectively). The water content in different measurements ranges from \SI{34}{ppm} to \SI{98}{ppm} (\SI{120}{ppm} to \SI{180}{ppm}, respectively) for the \arco{} (\neco{}, respectively) mixtures. The coloured error-bars represent the error due to the drift length uncertainty, while the black error bar represents the combined uncertainty of all other sources.}
\end{figure*}
\begin{figure*}
\centering%l u r o
\subfloat[Different $\textrm{CO}_2$ content in \arco{}]{\label{ch:immeas:fig:results:arco2}
\includegraphics[width=\columnwidth,trim = 63 0 53 30, clip=true]{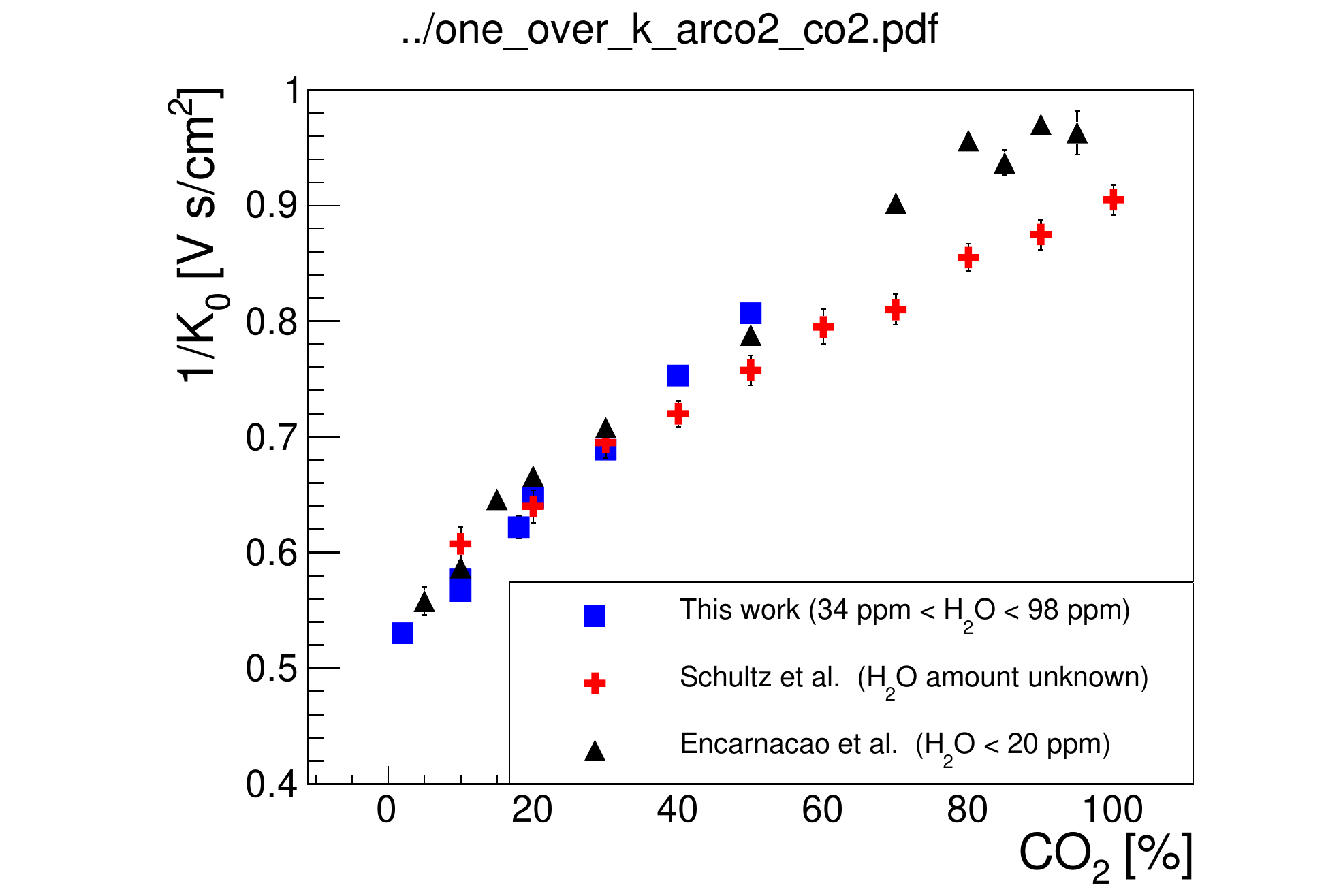}}
\subfloat[Different $\textrm{CO}_2$ content in \neco{}]{\label{ch:immeas:fig:results:neco2}
\includegraphics[width=\columnwidth,trim = 63 0 53 30, clip=true]{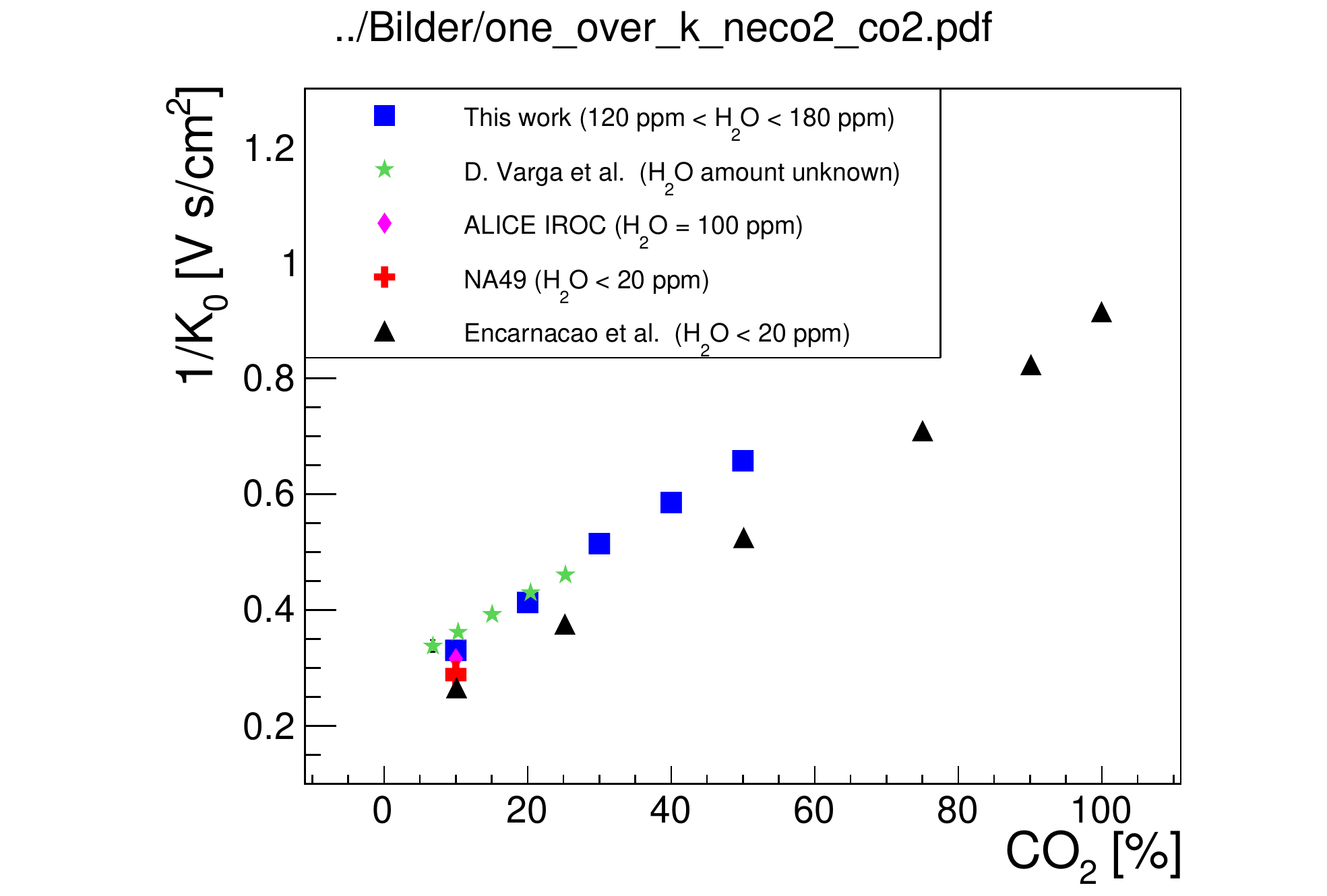}}
\caption{\label{ch:immeas:fig:results:nearco2}Ion mobilities for the different \arco{} mixtures \protect\subref{ch:immeas:fig:results:arco2} (respectively \neco{} mixtures \protect\subref{ch:immeas:fig:results:neco2}) result from the corresponding measurements in Figure \ref{ch:immeas:fig:result:ArCO2ED} (respectively \ref{ch:immeas:fig:result:NeCO2ED}). For comparison, results from other groups are shown: \textit{Schultz et al.} \textit{cf}. \cite{schultz1977mobilities}, \textit{Encarna{\c{c}}{\~a}o et al.} -- \arco{} \textit{cf}. \cite{encarnaccao2015experimental} and \neco{} \textit{cf}. \cite{1748-0221-11-05-P05005}, \textit{NA41} and \textit{ALICE IROC} \textit{cf}. \cite{kalkan2015cluster} and \textit{D. Varga et al.} \textit{cf}. \cite{dezso}.}
\end{figure*}
\begin{table*}
\centering
\begin{tabular}{l||c|c|c}
  &  $\textrm{Ar}$   &   $\textrm{CO}_{2}$ & $\chi^2/N_{\textrm{DF}}$ \\ \hline 
  &                  &                     & \\[-0.4cm]
$K_{0}$ $\left[\si{\centi\meter\squared\per\volt\per\second}\right]$  & $1.94\pm0.01$ & $1.10\pm0.01$ & 1.65\\
\end{tabular}\\
\vspace{0.4cm}
\begin{tabular}{l||c|c|c}
  &  $\textrm{Ne}$   &   $\textrm{CO}_{2}$ & $\chi^2/N_{\textrm{DF}}$ \\ \hline 
  &                  &                     & \\[-0.4cm]
$K_{0}$ $\left[\si{\centi\meter\squared\per\volt\per\second}\right]$ & $4.06\pm0.07$ & $1.09\pm0.01$ & 4.62\\
\end{tabular}
\caption{\label{ch:immeas:tab:nearco2mob}Fit results of a linear fit according to Equation \eqref{ch:immeas:eq:blancslawfit} are shown. The fit has been performed to our data displayed in Figures \ref{ch:immeas:fig:results:arco2} and \ref{ch:immeas:fig:results:neco2}. According to Blanc's law \cite{blanc1908pement}, the reduced ion mobilities here are the mobilities in pure argon, neon and carbon-dioxide for the ion species drifting in the \arco{} and \neco{} mixtures used in the measurements. $N_{\textrm{DF}}$ is the number of the degrees of freedom of the fit.}
\end{table*}
\begin{figure*}[ht]
\centering%l u r o
\subfloat[]{\label{ch:immeas:fig:results:neco2n2ED}
\includegraphics[width=\columnwidth,trim = 61 0 45 30, clip=true]{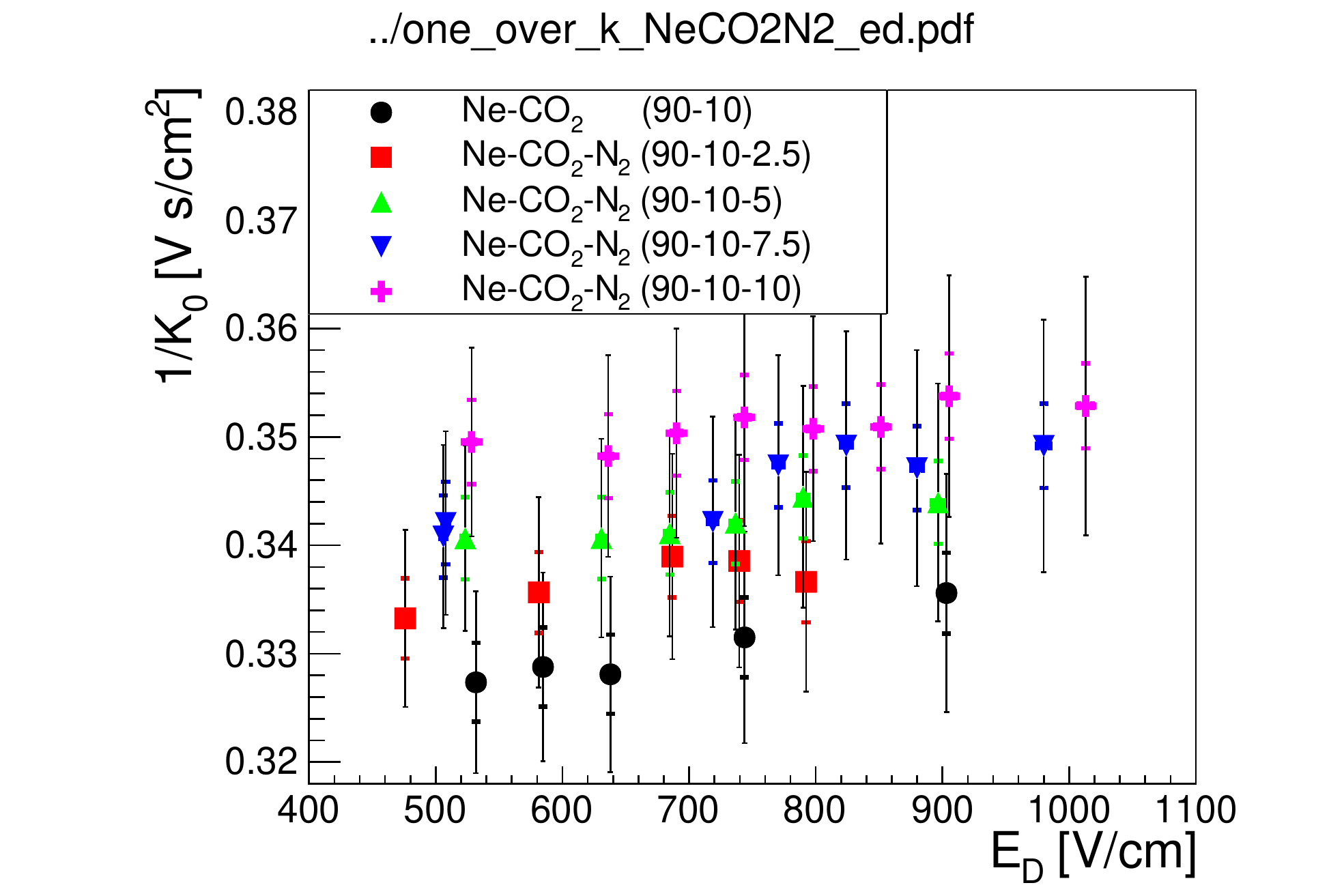}}
\subfloat[]{\label{ch:immeas:fig:results:neco2n2}
\includegraphics[width=\columnwidth,trim = 61 0 45 30, clip=true]{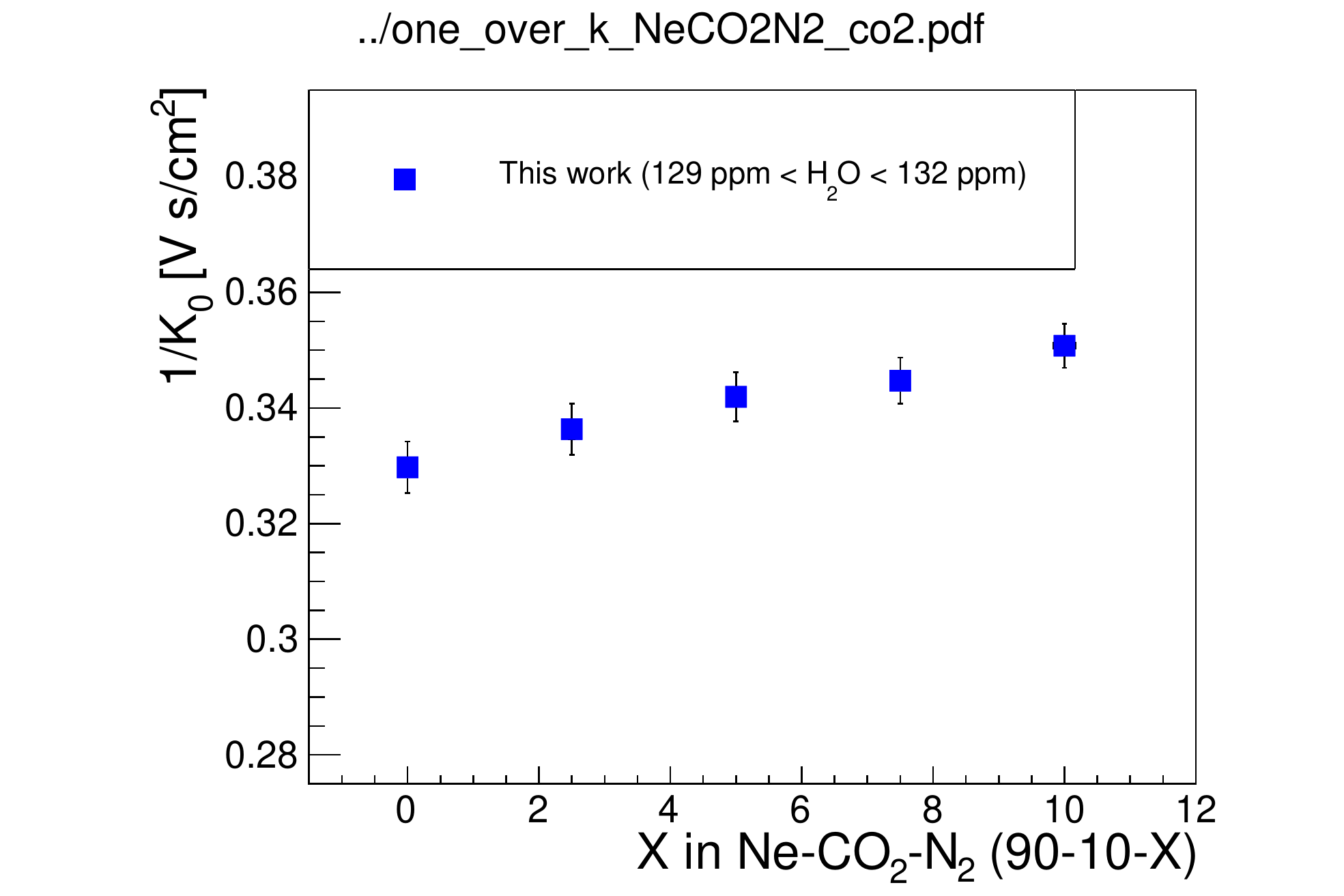}}
\caption{\label{ch:immeas:fig:results:neco2n2EDandMix}Measurements of $1/K_{0}$ for different amount of $\textrm{N}_{2}$ admixtures to a \necois{90-10} gas mixture. Figure \protect\subref{ch:immeas:fig:results:neco2n2ED} shows one over the reduced mobility for different \E{Drift} for different \baseline{} mixtures. The ratios (90-10-X) are not meant to be interpreted as \% ratios but mixing ratios. The same convention for the error bars as in Figure \ref{ch:immeas:fig:result:ArAndNeCO2ED} is used here. The points in \protect\subref{ch:immeas:fig:results:neco2n2} correspond to the average of each of these measurement series.}
\end{figure*}
\begin{figure*}
\centering%l u r o
\subfloat[\arcois{90-10}]{\label{ch:immeas:fig:results:arco2EDwater}
\includegraphics[width=\columnwidth,trim = 61 0 53 30, clip=true]{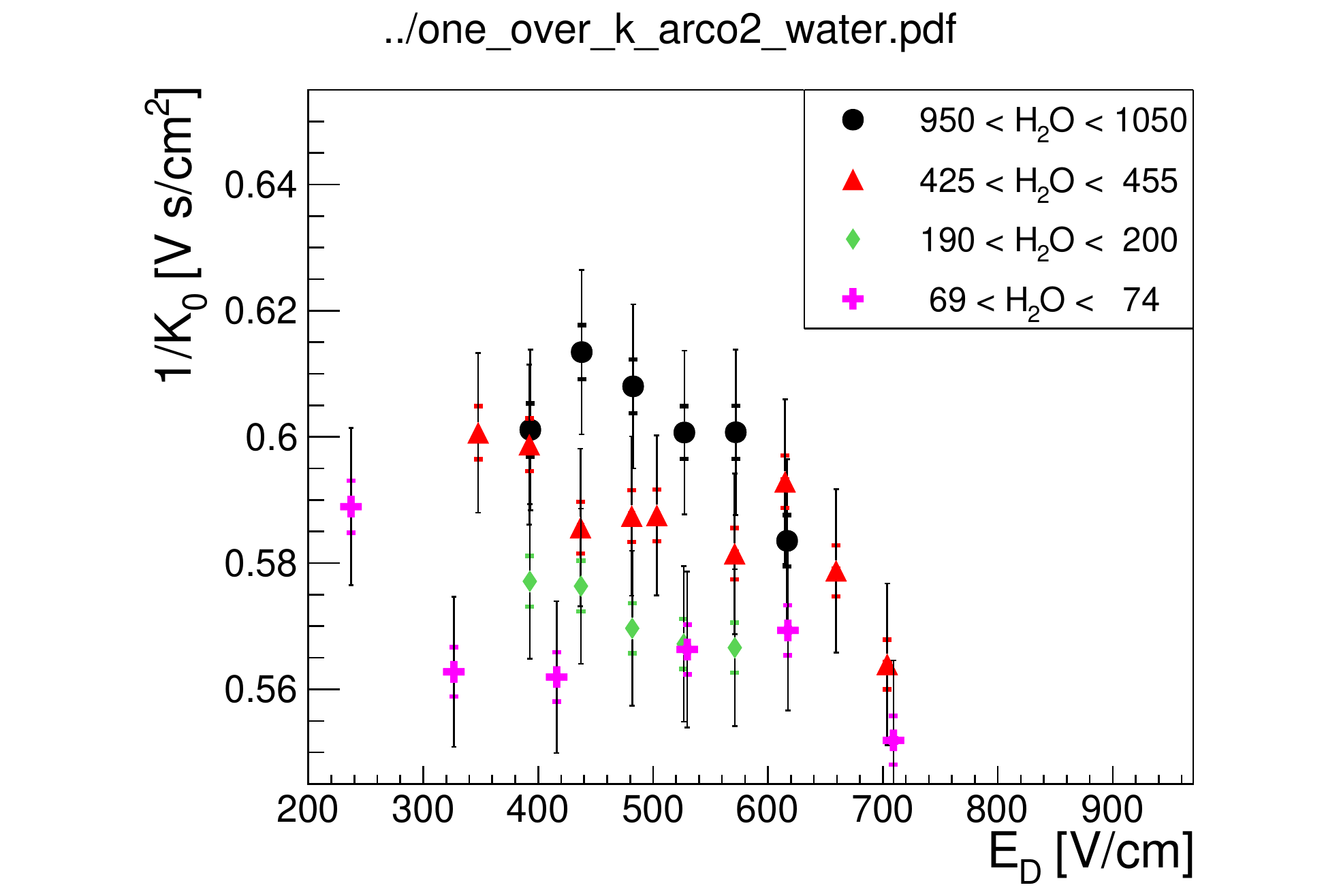}}
\subfloat[\necois{90-10}]{\label{ch:immeas:fig:results:neco2EDwater}
\includegraphics[width=\columnwidth,trim = 61 0 53 30, clip=true]{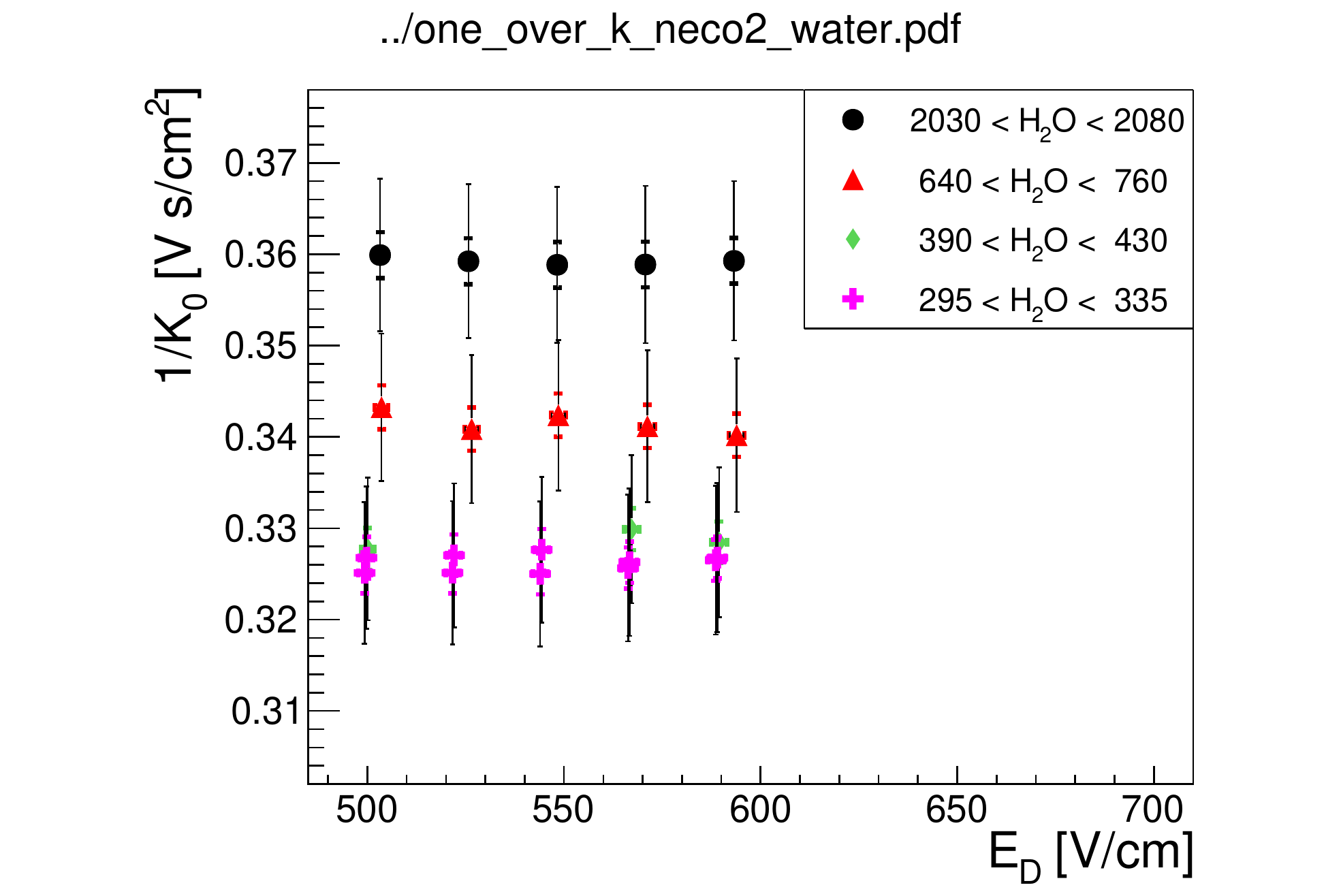}}
\caption{\label{ch:immeas:fig:results:arco2neco2EDH2O}The reduced ion mobility for varying water content in two gas mixtures. The same convention for the error bars as in Figures \ref{ch:immeas:fig:result:ArAndNeCO2ED} and \ref{ch:immeas:fig:results:neco2n2ED} applies here.}
\end{figure*}

\subsection{Mobility as function of $\mathbf{E_{\textrm{Drift}}}$}
\label{ch:immeas:subsec:koversusE}
Figure \ref{ch:immeas:fig:result:ArAndNeCO2ED} shows the inverse of the reduced mobility $K_{0}$ as a function of $\textrm{E}_{\textrm{Drift}}$ for different \arco{} and \neco{} mixtures. The drift field (\E{Drift}$/N$, respectively) ranges from \SI{200}{\volt\per\centi\meter} to \SI{1100}{\volt\per\centi\meter} (\SI{0.84}{Td} to \SI{4.6}{Td}, respectively\footnote{$N$ is the neutral particle density and $\si{Td}=\SI{1e-17}{\volt\centi\meter\squared}$.}) while the average gas pressure and temperature is $\SI{966(9)}{\milli\bar}$ and $293.5\pm\SI{1.5}{\kelvin}$, respectively. For these values of \E{Drift}, the (reduced) ion mobility of each mixture is found to be constant, which is the expected behaviour for low electric fields. (For high fields the mobility changes with the electric field with a $K\sim1/\sqrt{E}$ dependence \cite{blum2008particle}.) An electric field is considered as low, if the energy, gained by an ion from the field, is much smaller than the thermal energy (\textit{cf.} \cite{mcdaniel1973mobility}) \textit{i.e.} if
\begin{align}
\label{ch:immeas:eq:isitalowfield} \left(\frac{m_1}{m_2}+\frac{m_2}{m1}\right)e\textrm{E}l \ll k T \quad,
\end{align}
where the masses $m_{1}$ and $m_{2}$ have to be chosen according to the masses of the ion and its collision partner and $l$ is the mean free path between collisions. The mass factor in Equation \eqref{ch:immeas:eq:isitalowfield} will be $\geq2$, considering the masses of the atoms and molecules in the used gas mixtures. The inequality is thus not fulfilled ($l=\SI{100}{\nano\meter}$, \E{}$\ = \SI{1000}{\volt\per\centi\meter}$)\footnote{At $P=\SI{1013}{\milli\bar}$ and $T=\SI{273.15}{\kelvin}$ the mean free path between collisions is \SI{39.5}{\nano\meter}, \SI{63.2}{\nano\meter} and \SI{125}{\nano\meter} for pure $\textrm{CO}_2$, $\textrm{Ar}$ and $\textrm{Ne}$, respectively \cite{paths}.} at our gas conditions, however, the energy gained from the field is not much larger than $kT$. We therefore measure $K(\textrm{E}_{\textrm{Drift}})$ in a range where the transition from $K\sim\textrm{const}$ to $K\sim1/\sqrt{E}$ takes place and the observed constant $K_{0}^{\textrm{Mix}}(\textrm{E}_{\textrm{Drift}})$ are reasonable. 

\subsection{Blanc's law for $\textrm{Ar}$- and \neco{}}
\label{ch:immeas:subsec:blancarconeco}
Since the measured mobilities do not depend on the drift field, a weighted mean of the measurements is calculated, weighting each measurement with its uncertainty. The corresponding error of the weighted mean is calculated and added in quadrature with the systematic uncertainty on the drift length.\\ \indent
No indication for more than one drifting ion species is observed in the data, which fits the expectation that only one type of ($\textrm{CO}_2$ based cluster) ion drifts in the \arco{} and the \neco{} mixtures \cite{kalkan2015cluster}. Hence Blanc's law \cite{blanc1908pement} can be applied in order to extract the mobility of these ions in pure argon (respectively neon) and pure carbon-dioxide. To this end a fit of the type
\begin{align}
\frac{1}{K_{0}^{\textrm{Mix}}} = a + f_{\textrm{CO}_2}\left(b-a\right)
\label{ch:immeas:eq:blancslawfit}
\end{align}
is applied to the data in each panel of Figure \ref{ch:immeas:fig:results:nearco2}. The parameter $a$ and $b$ correspond to $1/K_{0}^{\textrm{Ar}}$ (respectively $1/K_{0}^{\textrm{Ne}}$) and $1/K_{0}^{\textrm{CO}_2}$, respectively, while $f_{\textrm{CO}_2}$ is the $\textrm{CO}_2$ fraction in the mixture. The superscript on the $K_0$s indicates the mobility of the drifting ion in the particular gas or mixture. The extrapolated mobility at \SI{100}{\%} $\textrm{CO}_2$, $K_{0}^{\textrm{CO}_2}$, is compatible, within errors, for both argon and neon mixtures (Table \ref{ch:immeas:tab:nearco2mob}). This may indicate that in both mixtures the same (cluster) ion species drifts. The results obtained by the presented method are compatible with other measurements \cite{schultz1977mobilities,encarnaccao2015experimental,kalkan2015cluster,dezso}. \\ \indent
We compare our results in Table \ref{ch:immeas:tab:nearco2mob} to results of ion mobility measurements for different argon\mbox{-,} neon\mbox{-,} nitrogen\mbox{-,} and carbon-dioxide-based (cluster) ions drifting in the respective pure gases \cite{Ellis1976,Ellis1978,Ellis1984,VIEHLAND199537}. Most of the mobilities included in these reviews are obtained at different $\textrm{E}/N$, $T_{\textrm{Meas}}$ or $P_{\textrm{Meas}}$ as compared to the gas conditions present during this work. In \cite{doi:10.1063/1.432024} a reduced mobility of $\SI{1.06(2)}{\centi\meter\squared\per\volt\per\second}$ is reported for a not identified carbon-dioxide based cluster ion, possible containing hydrogen, measured in $\textrm{CO}_2$ ($T_{\textrm{Meas}}=\SI{298}{\kelvin}$, $P_{\textrm{Meas}}=\SI{1016}{\milli\bar}$). This value is compatible with our $K_{0}^{\textrm{CO}_2}$, gained from the \arco{} and \neco{} measurements (Table \ref{ch:immeas:tab:nearco2mob}). Thus the comparison to the \cite{doi:10.1063/1.432024} data is a strong indication that the drifting ion species in \arco{} and \neco{} is a $\textrm{CO}_2$ based cluster ion as is suggested for these gas mixtures \cite{kalkan2015cluster}. A reduced mobility of $\sim\SI{1.1}{\centi\meter\squared\per\volt\per\second}$ for $\textrm{CO}_{2}^{+}\textrm{CO}_2$ drifting in carbon dioxide is reported in \cite{refId0,VIEHLAND199537} ($T_{\textrm{Meas}}=\SI{300}{\kelvin}$, $P_{\textrm{Meas}}=\SI{0.666}{\milli\bar}$). However, it has to be kept in mind that the ion clustering processes can differ for different gas pressure and temperature \cite{doi:10.1063/1.432024}, which makes an extrapolation between measurements at different $T_{\textrm{Meas}}$ and $P_{\textrm{Meas}}$ not significant.

\subsection{$\textrm{N}_2$ in \neco{} mixtures}
\label{ch:immeas:subsec:addingn2}
The addition of $\textrm{N}_2$ to the \necois{90-10} mixture is studied in order to determine $K_{0}$ of the baseline gas mixture of the upgraded ALICE TPC. Figure \ref{ch:immeas:fig:results:neco2n2ED} shows several $1/K_{0}^{\textrm{Ne}\textnormal{-}\textrm{CO}_{2}\textnormal{-}\textrm{N}_2}$ measurement series for \necois{90-10} and different amounts of nitrogen.\\ \indent
For Figure \ref{ch:immeas:fig:results:neco2n2} we average the $1/K_{0}^{\textrm{Ne}\textnormal{-}\textrm{CO}_2\textnormal{-}\textrm{N}_2}$ for each $\textrm{Ne}$-$\textrm{CO}_2$-$\textrm{N}_2$ mixture as done for the mixtures discussed in Section \ref{ch:immeas:subsec:blancarconeco}. Adding nitrogen to a \necois{90-10} gas mixture reduces the (reduced) ion mobility of the resulting mixture. The effect is however small, \textit{i.e.} changing from \SI{0}{\%} nitrogen to \SI{5}{\%} nitrogen, results in a reduction of $K_{0}^{\textrm{Mix}}$ of about \SI{3.3}{\%}.\\ \indent
A fit of Equation \eqref{ch:immeas:eq:blancslawfit} to the data with either $1/K_{0}^{\textrm{Ne}}$ or $1/K_{0}^{\textrm{CO}_2}$ fixed to the value in Table \ref{ch:immeas:tab:nearco2mob} yields the mobility of the drifting ion species in \baseline{} in \SI{100}{\%} nitrogen, $K_{0}^{\textrm{N}_{2}}=\SI{1.8(2)}{\centi\meter\squared\per\volt\per\second}$. The other free parameter -- either $K_{0}^{\textrm{Ne}}$ or $K_{0}^{\textrm{CO}_{2}}$ -- is about \SI{8}{\%} lower than the reduced mobility listed in Table \ref{ch:immeas:tab:nearco2mob}. This discrepancy together with the relatively high uncertainty on $K_{0}^{\textrm{N}_{2}}$ could be an indication that the drifting ion species in \baseline{} differs from the one in \neco{}. Nitrogen has a higher ionisation energy than carbon-dioxide, water or oxygen but a lower ionisation energy than neon \cite{NISTionisationenergies}. The charge transfer from $\textrm{Ne}$ ions to $\textrm{CO}_2$ molecules could perhaps be modified by the presence of $\textrm{N}_2$ molecules.\\ \indent
The reduced ion mobility of the baseline gas mixture for the future ALICE TPC \baseline{} (90-10-5) is \SI{2.92(4)}{\centi\meter\squared\per\volt\per\second}, measured at a water content of \SI{130(1)}{ppm}.

\subsection{Traces of $\textrm{H}_{2}\textrm{O}$ in \necois{90-10} and \arcois{90-10}}
\label{ch:immeas:subsec:resultswater}
\begin{figure}[h]
\centering%l u r o
\includegraphics[width=\columnwidth,trim = 63 0 53 30, clip=true]{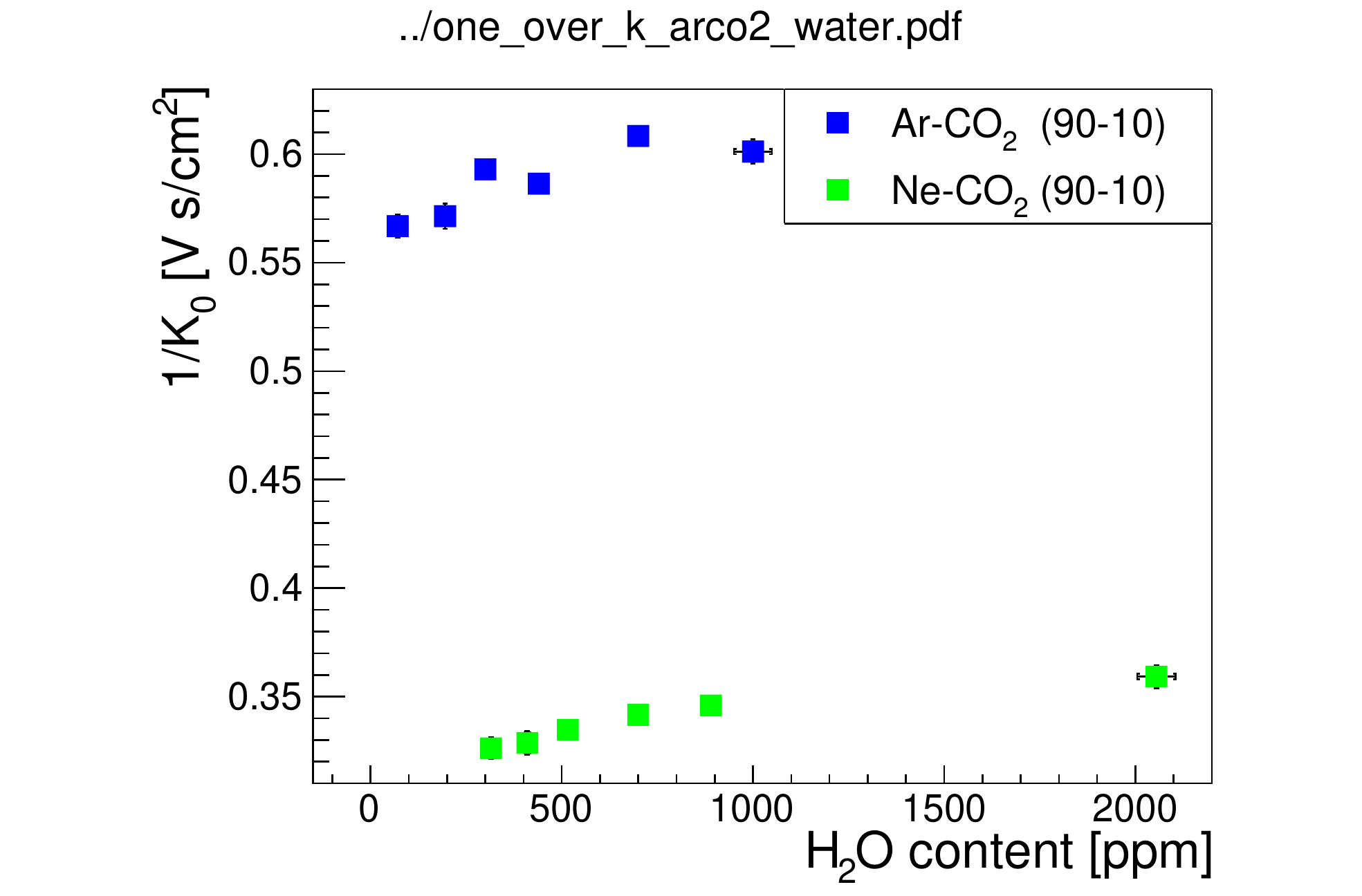}
\caption{\label{ch:immeas:fig:results:arco2neco2ED}The weighted mean of the data series in Figure \ref{ch:immeas:fig:results:arco2EDwater} (respectively Fig. \ref{ch:immeas:fig:results:neco2EDwater}) showing the change in mobility for varying water content in \arcois{90-10} and \necois{90-10}, respectively. In addition, results from measurements in \arcois{90-10} (\necois{90-10}, respectively) at a water content of \SI{700(5)}{ppm} and \SI{300(5)}{ppm} (\SI{890(20)}{ppm} and \SI{516(15)}{ppm}, respectively) are included as well in the figure, but are not shown in Figure \ref{ch:immeas:fig:results:arco2neco2EDH2O}. The horizontal error-bars do not represent an uncertainty, but indicate the range over which the water content is varied.}
\end{figure}

In order to have similar gas conditions as are present in the ALICE TPC most of our measurements are done with a water content around \SI{100}{ppm}. To study the effect of traces of $\textrm{H}_{2}\textrm{O}$ on the (reduced) ion mobility, the water content in \arcois{90-10} (\necois{90-10}, respectively) has been increased from about \SI{70}{ppm} to about \SI{1000}{ppm} (respectively from $\sim\SI{320}{ppm}$ to $\sim\SI{2050}{ppm}$) (Fig. \ref{ch:immeas:fig:results:arco2neco2EDH2O}). Water is introduced into the gas by inserting a given length of nylon tubing in the supply, since water diffuses through nylon walls.\\ \indent
In \arcois{90-10} (Fig. \ref{ch:immeas:fig:results:arco2EDwater}) no strong effect of water admixtures is visible. The points in the figure are a compilation of several measurements, partially with a large time between subsequent measurements. In case of the \necois{90-10} (Fig. \ref{ch:immeas:fig:results:neco2EDwater}), all measurements are done directly after each other in order to minimise other changes in the gas conditions. In this case, a clear decrease of the reduced mobility with increasing water content can be seen.\\ \indent
The weighted mean of each measurement series is again calculated (Fig. \ref{ch:immeas:fig:results:arco2neco2ED}). For the \arcois{90-10} mixture there is a decreasing trend for $K_{0}^{\textrm{Mix}}(\textrm{H}_{2}\textrm{O})$ as the water content is increased from \SI{72}{ppm} to about \SI{700}{ppm}. In \necois{90-10} an increase of the water content in the mixture from \SI{365}{ppm} to \SI{900}{ppm} leads to a clear reduction of the mobility by \SI{5.8}{\%}. A measurement series with even higher water content (\SI{2200}{ppm}) has been done, indicating that the decrease of mobility with increasing water content is more prominent at lower water content.\\ \indent
\neco{} mixtures are faster mixtures than \arco{} mixtures, with respect to the ion velocity. Therefore the impact of the $\textrm{H}_{2}\textrm{O}$ molecules is more drastic in the \neco{} case. For varying water content in a gas mixture there are no previous systematic studies in the literature, we are aware of, in order to compare to these measurements. They agree with the assumption that water lowers the ion mobility, as suggested in \cite{kalkan2015cluster}. In the same paper the authors observe a $\sim\SI{11}{\%}$ lower mobility in \necois{90-10} while comparing measurements done at a water content of less than \SI{20}{ppm} to a measurement with another detector using \SI{100}{ppm} of water. In case the mobility change as a function of $\textrm{H}_{2}\textrm{O}$ is this drastic at low water content, this change could explain the difference between the \neco{} results in \cite{1748-0221-11-05-P05005,kalkan2015cluster} and this work (Fig. \ref{ch:immeas:fig:results:neco2}). However, comparing between \cite{encarnaccao2015experimental} and our results (Fig. \ref{ch:immeas:fig:results:arco2}) shows that such an effect is not present in \arco{} mixtures.

\section{Summary}
\label{ch:immeas:sec:summary}
The drift time of ions through a drift gap was measured using two simultaneous recorded signals. The start signal is recorded on the electrode from which the ions initiate their drift. In order to determine the time of arrival of the ions at a wire-grid, defining the end of the drift gap, a novel method was developed. The induced signal over the full ion drift time is recorded. The inflection point of this signal is found to correspond to the ion's arrival time at the grid. From the difference of this two times the mobility is calculated. This has been done for different gas mixtures and for different drift field values in each gas mixture. The effect of water admixtures was examined as well.\\ \indent
The measurements were done for drift fields ranging between \SI{200}{\volt\per\centi\meter} and \SI{1100}{\volt\per\centi\meter} at ambient conditions, with a typical water content of about \SI{100}{ppm}.\\ \indent
Ion mobilities are found to be constant in the explored range of $\textrm{E}/P_{\textrm{Meas}}$. From fits of Blanc's law to the \arco{} data we found the reduced mobility of the drifting (cluster) ion in pure argon to be $\SI{1.94(1)}{\centi\meter\squared\per\volt\per\second}$ and in pure carbon-dioxide to be $\SI{1.10(1)}{\centi\meter\squared\per\volt\per\second}$. For similar fits to the \neco{} data we found the reduced mobility of the drifting (cluster) ion to be \SI{4.06(7)}{\centi\meter\squared\per\volt\per\second} in pure $\textrm{Ne}$ and $\SI{1.09(1)}{\centi\meter\squared\per\volt\per\second}$ in pure carbon-dioxide. The similarity of both mobilities in $\textrm{CO}_2$ suggest that the same ion drifts in the argon- as well as the neon-based gas mixture. From a similar analysis of nitrogen admixtures to \necois{90-10} we found a value of $\SI{1.8(2)}{\centi\meter\squared\per\volt\per\second}$ for the mobility of the drifting ion in pure nitrogen. It remains to show if the same ion species drifts in \baseline{} than in \neco{}. Admixtures of $\textrm{N}_2$ reduced the mobility as compared to pure \necois{90-10} by $\sim\SI{3.3}{\%}$ for an increase of the nitrogen content by \SI{5}{\%}. The reduced mobility of the baseline gas mixture of the upgraded ALICE time projection chamber, \baseline{} (90-10-5) was found to be $\SI{2.92(4)}{\centi\meter\squared\per\volt\per\second}$. This value was measured with a water content of $\SI{130}{ppm}\pm\SI{1}{ppm}$ in the gas mixture.\\ \indent
Furthermore the change of the ion mobility induced by the water content in the gas was examined using \arcois{90-10} and \necois{90-10} mixtures. In case of \arcois{90-10} a slight decrease of the mobility was observed as the water content is increases from \SI{72}{ppm} to \SI{700}{ppm}. While increasing the water content in the \necois{90-10} mixture from \SI{365}{ppm} to \SI{900}{ppm} the (reduced) mobility decreased by \SI{5.8}{\%}. The effect seemed to level off for a further increase of the $\textrm{H}_{2}\textrm{O}$ content.

\section{Acknowledgements}
The authors thank Dr D. Vranic for his contribution to assembling the experimental set-up and for fruitful discussions during the realisation of this work.

\section*{References}

\bibliography{im-paper}

\end{document}